\documentclass[pra,twocolumn]{revtex4-1}
\usepackage{xcolor}
\usepackage{graphicx}
\usepackage{amssymb, amsmath, amsthm}
\usepackage{bbold}
\usepackage{dsfont}
\usepackage[percent]{overpic}

\usepackage[colorlinks]{hyperref}

\usepackage[authormarkup=none]{changes}
\definechangesauthor[name={a}, color=orange]{a} 

\def \diracspacing {0.7pt}
\newcommand{\bra}[1]{\langle #1 \hspace{\diracspacing} |} 
\newcommand{\ket}[1]{| \hspace{\diracspacing} #1 \rangle} 
\newcommand{\ketbra}[2]{| \hspace{\diracspacing} #1 \rangle \langle #2 \hspace{\diracspacing} |} 
\newcommand{\avg}[1]{\langle #1 \rangle}

\newcommand{\dg}{\dagger}

\newcommand{\id}{\hat{\mathds{1}}}
\newcommand{\rmd}{\mathrm{d}}

\begin{document}
	
\author{Alexandre Roulet}
\affiliation{Department of Physics, University of Basel, Klingelbergstrasse 82, CH-4056 Basel, Switzerland}

\title{Revealing the work cost of generalized thermal baths}
\begin{abstract}
We derive the work cost of using generalized thermal baths from the physical equivalence of quantum mechanics under unitary transformations. We demonstrate our method by considering a qubit extracting work from a single bath to amplify a cavity field. There, we find that only half of the work investment is converted into useful output, the rest being wasted as heat. These findings establish the method as a promising tool for studying quantum resources within the framework of classical thermodynamics.
\end{abstract}
\maketitle
\emph{Introduction.--} 
Thermodynamics governs the task of converting heat into useful work. In the classical formulation of the theory, the source of heat is a thermal bath, which corresponds to the available source of disordered energy from which pioneers aimed to manufacture ice~\cite{carre60} and set trains into motion~\cite{carnot24}.

In recent years, quantum thermodynamics has emerged as a promising counterpart for heat machines operating in the quantum regime. A natural motivation for this line of research is the unprecedented miniaturization of devices to the scale of a few atoms, where they inevitably leave the realm of classical physics. An even greater promise is the search for quantum effects which, when properly harnessed, would grant access to a more efficient use of energy as compared to what classical thermodynamics allows.

A significant step has been made in this direction by using generalized thermal baths as quantum fuel. These resources equilibrate the working medium they power to a unitarily-transformed thermal state. Theoretical predictions and experimental realizations have demonstrated that they allow to outperform the Carnot bound formulated in terms of the underlying thermal state~\cite{rossnagel14,klaers17}, to induce an absorption refrigerator to operate for temperatures that would lead to heating~\cite{correa14,ions17}, or even to extract work from a single bath~\cite{scully03,klaers17}. While these achievements would appear at first sight to violate the laws of thermodynamics, it is crucial to note that the present studies have all been carried out without accounting for the overhead expense of working with generalized thermal baths~\cite{niedenzu18}. The question addressed, and positively answered, was thus whether such resources would be of any use, were they to be freely available.

Coming back to the primary aim of a better heat conversion with quantum machines, there remains the fundamental question of whether these resources are worth the money. In this work, we identify the work cost of employing any generalized thermal bath by exploiting the physical equivalence of quantum mechanics under unitary transforms. In other words, we perform a change of coordinates that reveals the work that is to be supplied on top of the heat exchange to engineer the coupling to the out-of-equilibrium bath of interest. The method is illustrated on an elementary heat engine that amplifies the field of a cavity out of a single bath. There, while the possibility of extracting work without the need for a second bath seems to provide a clear advantage, we find that the associated work investment actually exceeds the achieved amplification, with half of it being wasted by the heat machine.
\emph{Unitary equivalence.--} The existence of a unitary transformation relating two seemingly different theories is the essence of the Stone-von Neumann theorem, which established that wave and matrix mechanics are equivalent formulations of quantum mechanics~\cite{rosenberg04}. There the unitary is the evolution operator which implements a time translation, providing the change of coordinates that relates the representations to one another. Here, for an arbitrary working medium coupled to a generalized thermal bath, we aim to identify the representation in which the bath is thermal and thus only exchanges heat. Any additional energy exchange of hamiltonian origin will then be readily accounted for as work.

A generic working medium consists of an isolated quantum system with free Hamiltonian $\hat{H}_\text{sys}$. When powered by a weakly-coupled thermal bath at temperature $T$, the state of the system thermalizes according to the master equation $\rmd \hat{\rho}/\rmd t=-i[\hat{H}_\text{sys},\hat{\rho}]/\hbar+\mathcal{L}_T [\hat{\rho}]$, where $\mathcal{L}_T$ is a superoperator of Lindblad form representing the dissipative coupling~\cite{gardiner85}. Specifically, its fixed point $\mathcal{L}_T [\hat{\rho}_T]=0$ is the thermal state $\hat{\rho}_T=e^{-\hat{H}_\text{sys}/k_B T}/Z$, with the partition function $Z=\text{Tr}[e^{-\hat{H}_\text{sys}/k_B T}]$ ensuring that the state is normalized. Now a generalized thermal bath is a variant of this situation where the system equilibriates to $\hat{\rho}_{U,T}=\hat{U}\hat{\rho}_T\hat{U}^\dg$ with $\hat{U}$ any time-independent unitary of interest for the task at hand. The form of the dissipative coupling is thus obtained by applying the unitary transformation $\hat{\rho}\to\hat{U}\hat{\rho}\hat{U}^\dg$, yielding $\mathcal{L}_{U,T} [\hat{\rho}]=\hat{U}\mathcal{L}_{T} [\hat{U}^\dg\hat{\rho}\hat{U}]\hat{U}^\dg$. A squeezed or a displaced thermal bath would for instance correspond to $\hat{U}$ being respectively the squeezing or displacement operator~\cite{ekert90}.

Before we proceed any further, note that this equilibriation takes place in the interaction picture. Indeed, if we were to simply replace $\mathcal{L}_{T} [\hat{\rho}]$ by $\mathcal{L}_{U,T} [\hat{\rho}]$ in the laboratory frame, the free Hamiltonian $\hat{H}_\text{sys}$ would wash out any attempt of the weakly-coupled bath to generate coherence in the basis of the energy eigenstates~\cite{gardiner04}. The coupling of our working medium to a generalized thermal bath is thus described by the master equation $\rmd \hat{\rho}/\rmd t=-i[\hat{H}_\text{sys},\hat{\rho}]/\hbar+\mathcal{L}^{(t)}_{U,T} [\hat{\rho}]$ where $\mathcal{L}^{(t)}_{U,T} [\hat{\rho}]=\hat{U}^{(t)}\mathcal{L}_{T} [\hat{U}^{(t)\dg}\hat{\rho}\hat{U}^{(t)}]\hat{U}^{(t)\dg}$ and the unitary is given by
\begin{equation}\label{eq:Ut}
	\hat{U}^{(t)}=e^{-i\hat{H}_\text{sys} t/\hbar}\hat{U}e^{i\hat{H}_\text{sys} t/\hbar} .
\end{equation}
It is in general time dependent, unless $\hat{U}$ is a symmetry of the system $[\hat{U},\hat{H}_\text{sys}]=0$ and the resource is then trivially thermal $\hat{\rho}_{U,T}=\hat{\rho}_T$.

\begin{figure}
	\centering
	\includegraphics[width=\columnwidth]{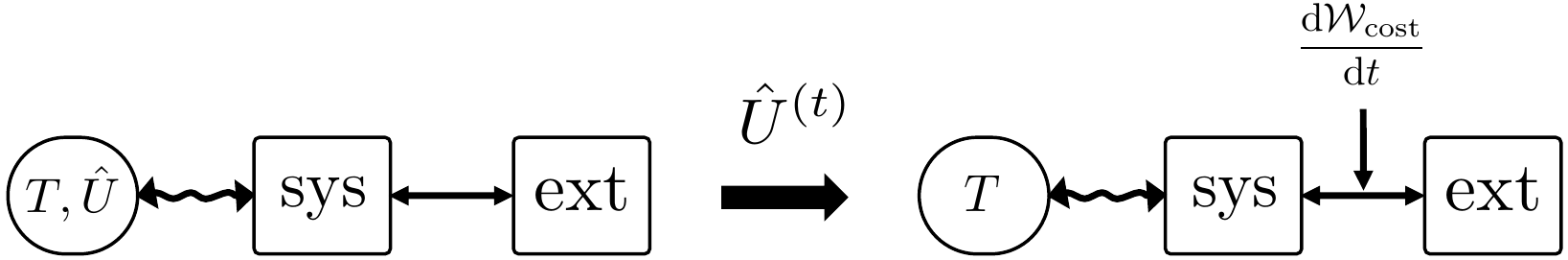}
\caption{\label{fig:diag} Diagram of the unitary transformation~\eqref{eq:master}. The generalized thermal bath is mapped to a thermal bath at the original temperature $T$. The latter exchanges only heat with the working medium, while an additional work source is required to mediate the coupling to the external system.}
\end{figure}

\begin{figure}
	\vspace{0.3cm}
	\centering
	\begin{overpic}[width=0.45\columnwidth]{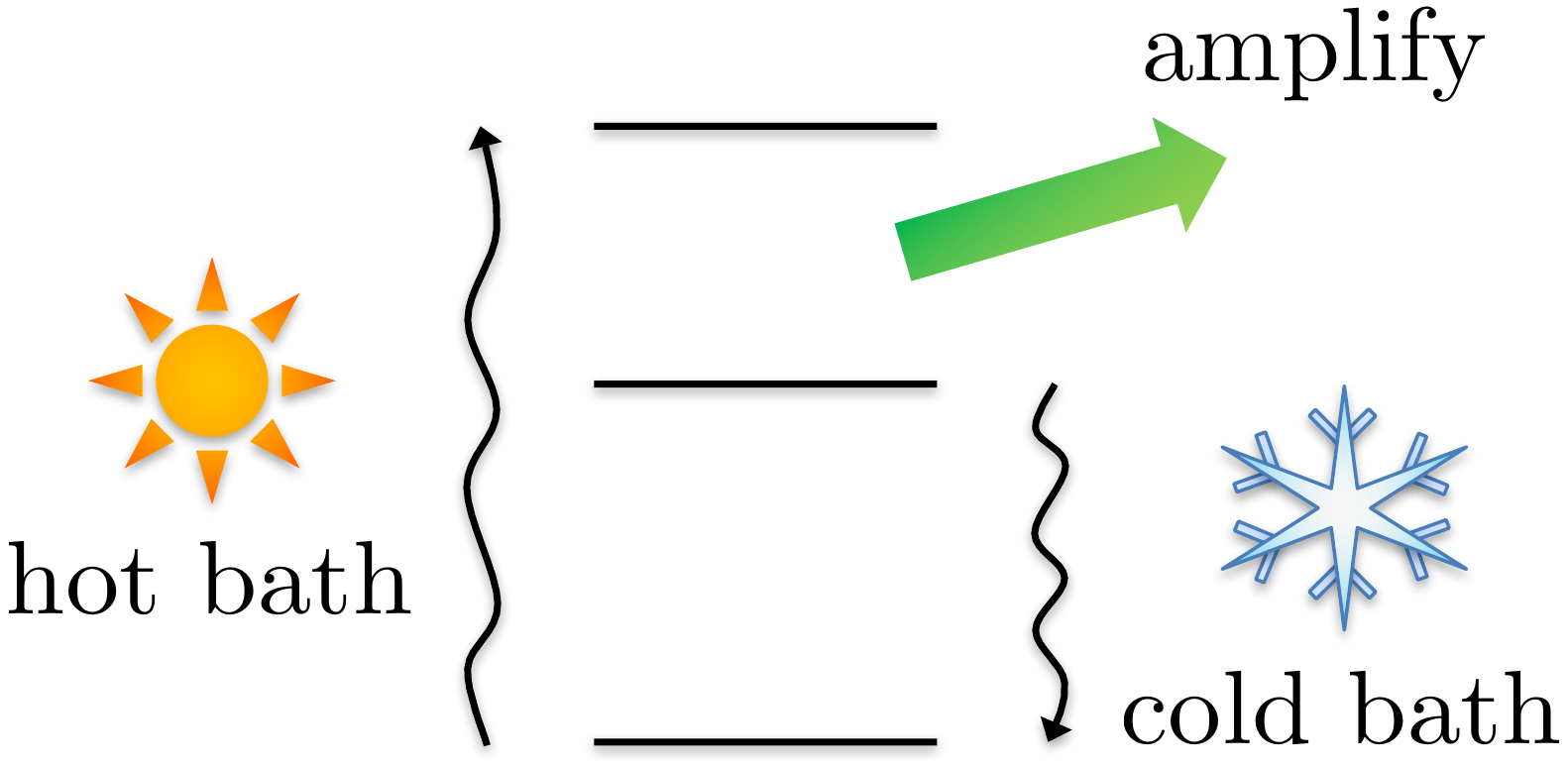}
		\put (0,47) {(a)}
	\end{overpic}\qquad
	\begin{overpic}[width=0.47\columnwidth]{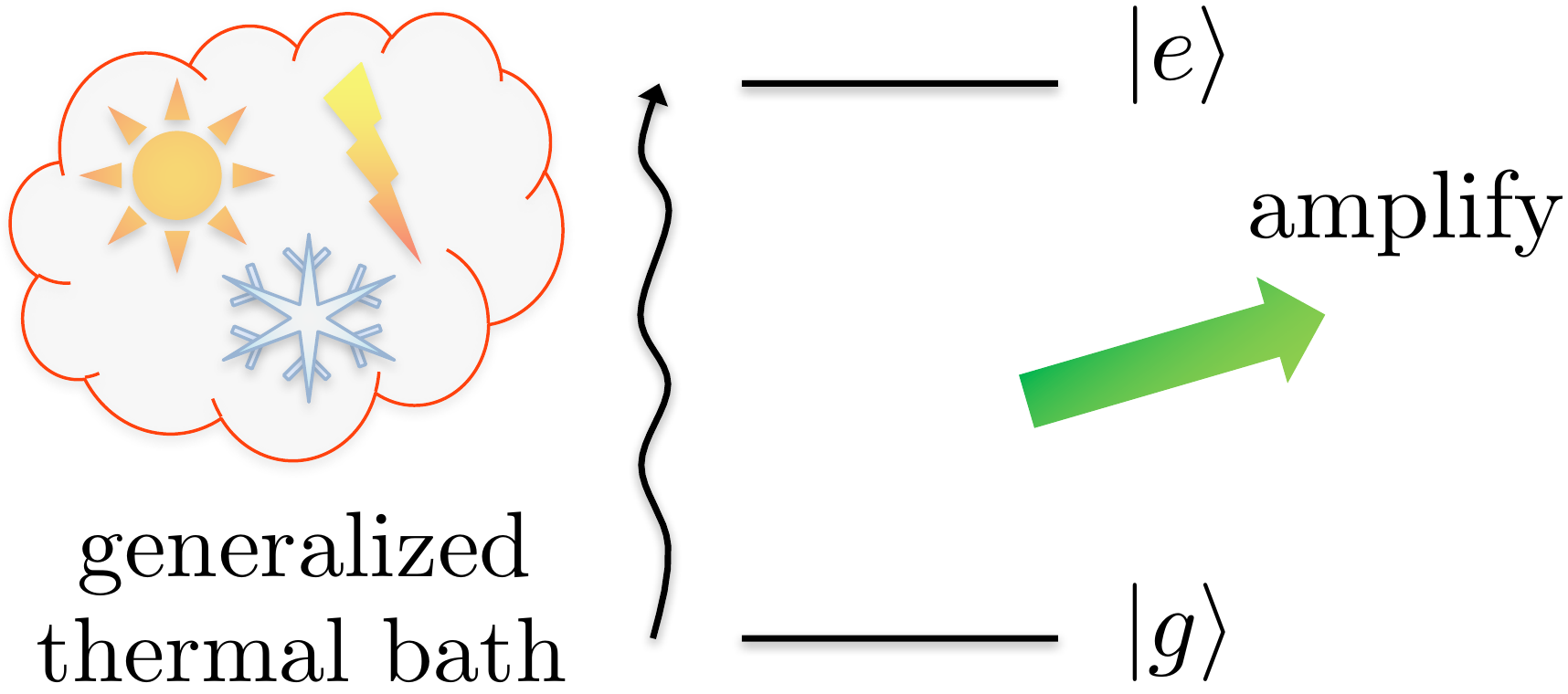}
		\put (0,47) {(b)}
	\end{overpic}
\caption{\label{fig:amp} A quantum engine converting energy into the amplification of a light field. (a) In the original formulation, the engine draws heat from a hot bath, emits into the light field and rejects extra heat into a cold bath, satisfying Carnot's principle. (b) Coupling the transition of interest to a generalized thermal bath, we aim for the engine to extract work from a single resource.}
\end{figure}

Equation~\eqref{eq:Ut} will be the main ingredient for obtaining the work cost we are after.
However, for now, it actually establishes that no extra cost is to be paid if one is solely aiming for a working medium being coupled to a generalized thermal bath. Indeed, it is then sufficient to rely on a standard thermal bath providing only heat and to make use of the unitary equivalence under the change of coordinates $\hat{\rho}\to\hat{U}^{(t)}\hat{\rho}\hat{U}^{(t)\dg}$. This first result highlights that we are not characterizing the process of creating and maintaining the out-of-equilibrium infinite-dimensional bath itself. Instead, our method assesses the operational cost of coupling to such a bath given that we have a thermal bath at hand. This is similar to the experimental reality where the dissipative coupling is engineered by acting onto the system, typically with the help of a laser drive~\cite{rugar91,poyatos96}, without modifying the bath.
\emph{Work cost.--} Surely generalized thermal baths have been shown to provide a competitive advantage when used to power quantum machines. For instance, a squeezed thermal bath employed in turn of a thermal bath increases the average energy of the working medium without changing its entropy~\cite{kim89}, improving the performance~\cite{rossnagel14}. Even more compelling is the possibility of extracting work from a single bath~\cite{scully03}. The key point here is that the benefits of the transformed bath are reaped when the working medium is used to extract work, that is when it interacts via some Hamiltonian $\hat{V}$ with an external system onto which it exerts work. In that case, the unitary equivalence to a thermal bath reads
\begin{widetext}
\begin{equation}\label{eq:master}
	\frac{\rmd \hat{\rho}}{\rmd t}=-\frac{i}{\hbar}\left[\hat{H}_\text{sys}+\hat{H}_\text{ext}+\hat{V},\hat{\rho}\right]+\mathcal{L}^{(t)}_{U,T} \left[\hat{\rho}\right] \quad\to\quad \frac{\rmd \hat{\rho}}{\rmd t}=-\frac{i}{\hbar}\left[\hat{H}_\text{sys}+\hat{H}_\text{ext}+\hat{U}^{(t)\dg}\hat{V}\hat{U}^{(t)},\hat{\rho}\right]+\mathcal{L}_T \left[\hat{\rho}\right] ,
\end{equation}
\end{widetext}
where $\hat{H}_\text{ext}$ is the free Hamiltonian of the external system. As illustrated in Fig.\,\ref{fig:diag}, this implies that for the change of coordinates to map the two physical situations, one is required to supply the work power
\begin{equation}\label{eq:cost}
	\frac{\rmd \mathcal{W}_\text{cost}}{\rmd t}=\text{Tr}\left[\partial_t\left(\hat{U}^{(t)\dg}\hat{V}\hat{U}^{(t)}\right) \hat{\rho}\right] ,
\end{equation}
on top of the thermal channel.
\emph{Amplifier.--}
The work cost identified in Eq.~\eqref{eq:cost} is derived for any generalized thermal bath powering any quantum machine. We will now focus on a concrete example to demonstrate how it can be easily computed to assess the return on work investment. Specifically, we consider the task of amplifying light, for which an elementary quantum heat machine was devised 60 years ago~\cite{scovil59}. It consists of a three-level system shown in Fig.\,\ref{fig:amp}(a) with each transition being respectively coupled to a hot bath, a cold bath, and the light mode to be amplified. Remarkably, this machine can reach Carnot efficiency, which is the maximum classical thermodynamics allows in terms of heat conversion. Yet, we will modernize its design to run on a single generalized thermal bath, aiming for a performance beyond the capabilities of a classical heat machine.
\begin{figure*}
	\centering
	\begin{overpic}[width=0.63\columnwidth]{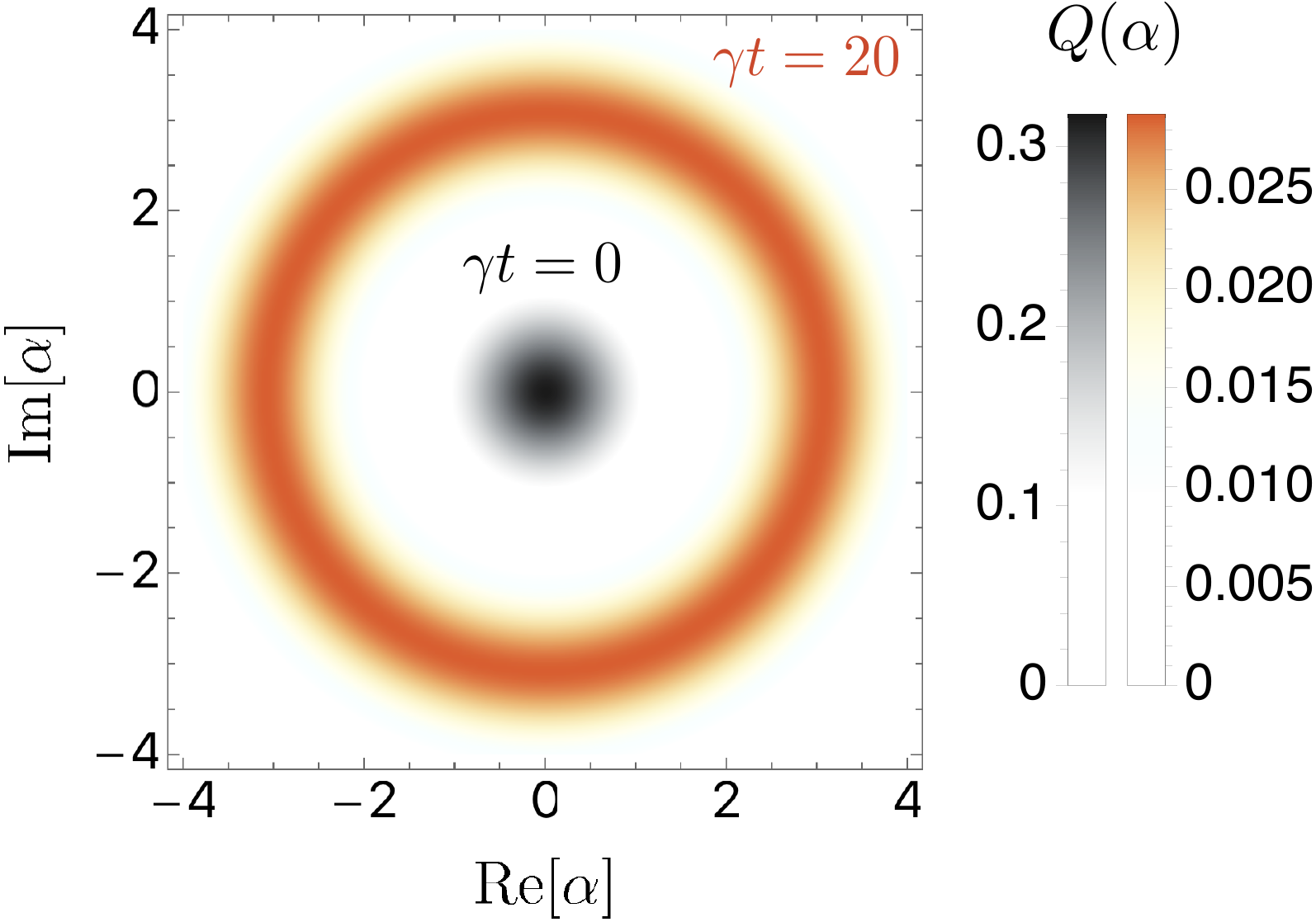}
		\put (0,70) {(a)}
	\end{overpic}\qquad
	\begin{overpic}[width=0.57\columnwidth]{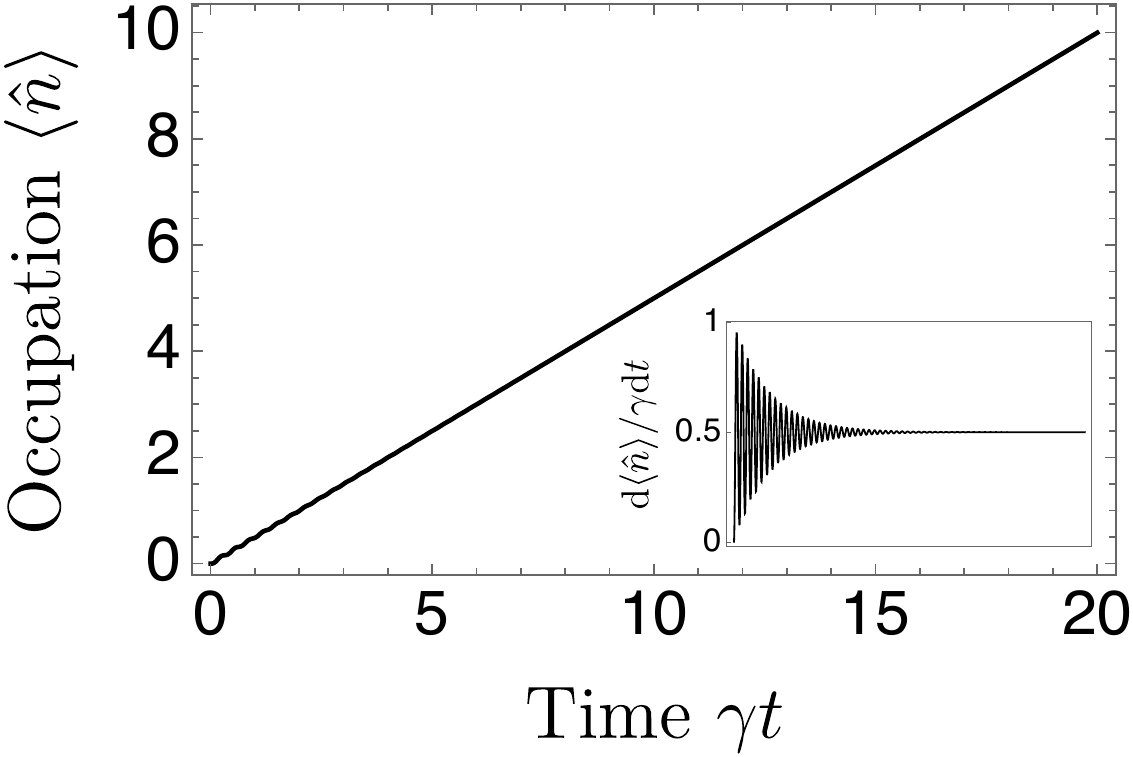}
		\put (0,74) {(b)}
	\end{overpic}\qquad
	\begin{overpic}[width=0.6\columnwidth]{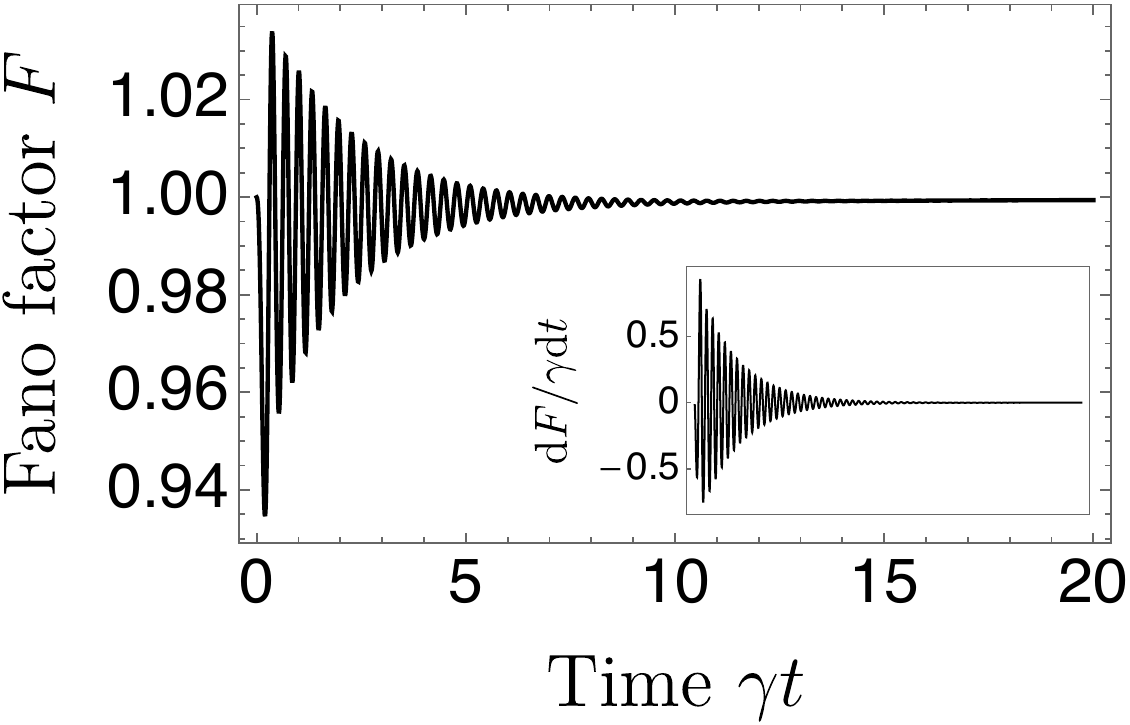}
		\put (0,70) {(c)}
	\end{overpic}
\caption{\label{fig:cav} The amplifier in operation for $g/\gamma=10$. (a) The Husimi Q representation of the light field at $\gamma t=0$ (black) and $\gamma t=20$ (orange). The engine displaces the state away from the origin without any phase preference. (b) The average number of photons in the light field $\avg{\hat{n}}$ increases linearly as a function of time $\gamma t$. (c) The Fano factor $F$ stabilizes to unity, corresponding to Poissonian statistics. The insets in (b) and (c) show the time derivative of the respective quantities, indicating that the light field reaches a steady-state regime.}
\end{figure*}
As illustrated in Fig.\,\ref{fig:amp}(b), we consider a working medium with a single transition at frequency $\omega$, \emph{i.e.} a qubit with free Hamiltonian $\hat{H}_\text{sys}=\hbar\omega\hat{\sigma}_z/2$ where $\hat{\sigma}_z$ is a Pauli matrix. In the original engine design, the role of the thermal baths was to invert the population of the transition coupled to the light field~\cite{geva96,tannor07}, thereby favoring emission (amplification) over absorption (attenuation). Here, we achieve this inversion by employing a \emph{flipped vacuum bath}, which corresponds to a vacuum bath $T=0$ transformed according to the Pauli matrix $\hat{U}=\hat{\sigma}_x$. When put into contact with the qubit, it equilibriates to the excited state $\hat{\rho}_{\sigma_x,0}=\ketbra{e}{e}$ while any thermal bath would lead to at least 50\% of the population in the ground state $\ketbra{g}{g}$. To complete the description of the quantum machine, we model the light field as a resonant bosonic mode, \emph{e.g.} a cavity field, with free Hamiltonian $\hat{H}_\text{ext}=\hbar\omega(\hat{a}^\dg \hat{a}+1/2)$ where $\hat{a}$ is the annihilation operator. It interacts with the working medium via $\hat{V}=i\hbar g(\hat{\sigma}_+\hat{a}-\hat{a}^\dg\hat{\sigma}_-)$, which corresponds to an exchange of excitations.

Now that the machine is assembled, we first test numerically whether light amplification takes place at all by initializing to the ground state with an empty cavity $\ket{g}\otimes\ket{0}$ and letting the engine run according to the master equation on the left of Eq.\,\eqref{eq:master}. In Fig.\,\ref{fig:cav}(a), the phase space distribution of the light field is shown employing the Husimi Q representation $Q(\alpha)=\bra{\alpha}\hat{\rho}_\text{ext}\ket{\alpha}/\pi$ where $\hat{\rho}_\text{ext}=\text{Tr}_\text{sys}[\hat{\rho}]$ is obtained by tracing out the working medium and $\ket{\alpha}$ is a coherent state of amplitude $\alpha$. The ring obtained from the initial centered vacuum state is a conclusive signature of the desired amplification, corresponding to the state being displaced in phase space. As the energy is extracted from a generalized thermal bath, the system does not possess any phase reference and the resulting light field is thus rotationally symmetric. The statistics of the light field is further characterized in (b) and (c), where the occupation is found to increase linearly in time with a Fano factor $F=(\avg{\hat{n}^2}-\avg{\hat{n}}^2)/\avg{\hat{n}}$ reaching unity, yielding the Poissonian statistics of a coherent state. 

The numerical results thus demonstrate that the working medium successfully leverages the single generalized thermal bath to steadily amplify the light field. In fact, as the master equation contains all the information on the engine's dynamics, we have in principle access to the analytical description of the amplification process. However, as for the original engine of Fig.\,\ref{fig:amp}(a)~\cite{tannor06} and despite the working medium being simplified to a qubit, the non-linearity of the working medium combined with the infinite Hilbert space of the bosonic mode complicate any attempt to solve the dynamics analytically. We thus impose Poissonian statistics as an ansatz and find that this is sufficient to characterize the steady-state operation of the engine. In particular, the field occupation $\hat{n}=\hat{a}^\dg\hat{a}$ is found to increase at a constant rate set by the repumping of the qubit by the bath $\rmd \avg{\hat{n}}_\text{ss}/\rmd t = \gamma/2$, in agreement with the inset of Fig.\,\ref{fig:cav}(b), while the qubit occupation equilibriates to $\avg{\hat{\sigma}_z}_\text{ss}=0$. This yields the following energy transfers in steady-state
\begin{equation}\label{eq:ss}
	\frac{\rmd \avg{\hat{H}_\text{ext}}_\text{ss}}{\rmd t}= \frac{\hbar \omega \gamma}{2}\qquad\text{and}\qquad \frac{\rmd \avg{\hat{H}_\text{sys}}_\text{ss}}{\rmd t}=0 ,
\end{equation}
where the second equation indicates that the qubit converts all the energy it takes from the bath into light amplification.

\emph{Work cost.--} We have thus identified an elementary quantum machine capable of extracting useful ordered energy from a single generalized thermal bath. This finding is in agreement with previous studies~\cite{scully03,klaers17}, and validates the usefulness of these quantum resources were they to be sold at the service station for the same price as their thermal counterpart. The method developed here allows us to take an important step further and assess whether the use of the vacuum flipped bath is thermodynamically favorable, and possibly even beyond the reach of any classical heat machine.

We first transform to a unitarily-equivalent representation where the bath is thermal, corresponding to the free resource of thermodynamics. This is done via the unitary given in Eq.\,\eqref{eq:Ut}, which reads for the present engine
\begin{equation}
	\hat{U}^{(t)}=e^{-i\omega t\hat{\sigma}_z/2}\hat{\sigma}_x e^{i\omega t\hat{\sigma}_z/2} .
\end{equation}
The change of coordinates implemented by this transformation is a rotation at the qubit frequency followed by a spin flip and the same rotation backwards. Importantly, the external system which receives the output of the machine cannot distinguish between the two representations, with $[\hat{U}^{(t)}\otimes\id,\id\otimes\hat{\mathcal{O}}]=0$ for any operator $\hat{\mathcal{O}}$ acting on the light field. However, following the recipe provided in Eq.\,\eqref{eq:master}, we find that the coupling of the amplifier to the light field is now mediated by an external coherent pump $\hat{U}^{(t)\dg}\hat{V}\hat{U}^{(t)}= i\hbar g(\hat{\sigma}_-\hat{a}e^{2i\omega t}-\hat{a}^\dg\hat{\sigma}_+e^{-2i\omega t})$ on top of the thermal channel, such that Eq.~\eqref{eq:cost} yields
\begin{equation}
	\frac{\rmd \mathcal{W}_\text{cost}}{\rmd t}=2\frac{\rmd \avg{\hat{H}_\text{ext}}}{\rmd t} \xrightarrow{ss}\hbar \omega \gamma ,
\end{equation}
where the last term corresponds to the steady-state limit.

\begin{figure}
	\centering
	\includegraphics[width=\columnwidth]{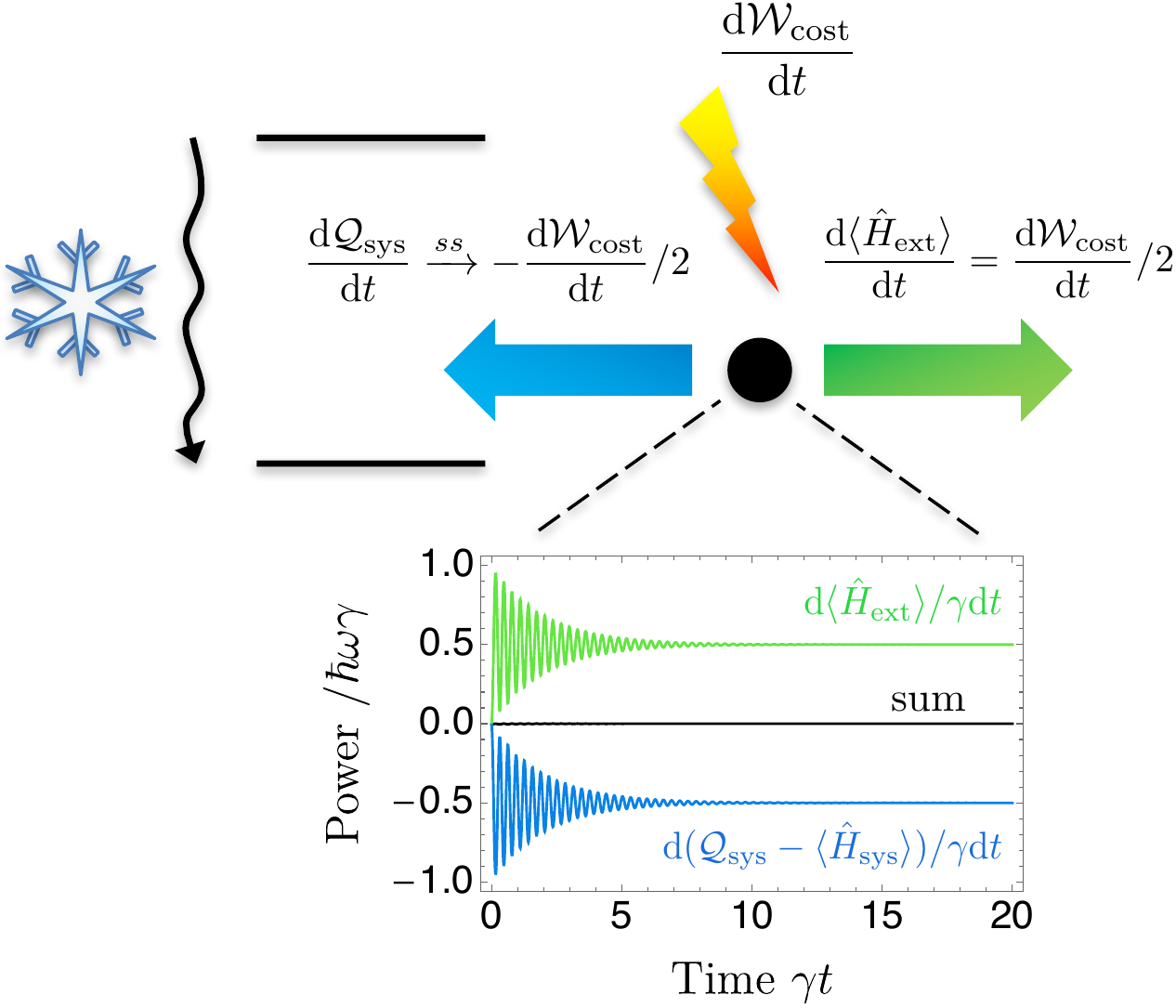}
\caption{\label{fig:pow} Half of the work investment $\rmd \mathcal{W}_\text{cost}/\rmd t$ is used to amplify the light field (green) while the other half is lost to the qubit (blue). Once the engine enters the steady-state regime, this latter part is exclusively dissipated as heat to the vacuum bath. The graph shows the energy flows as a function of time $\gamma t$ for the same parameters as in Fig.\,\ref{fig:cav}. Their sum (black) vanishes, satisfying the first law of thermodynamics, and the steady-state values agree with Eqs.\,\eqref{eq:ss} and \eqref{eq:heat}.}
\end{figure}
That the task performed by the quantum machine relies on a work investment is not necessarily a downside, as the benefits in terms of heat extraction may still very well pay off. Regrettably, the cost here is twice the energy received by the light field. The origin of this rather inefficient work-to-work conversion is manifest in this representation: half of the drive's input energy is used to excite the qubit via $\hat{a}^\dg\hat{\sigma}_+e^{-2i\omega t}$. The qubit then relaxes to the ground state $\ket{g}$ by releasing the excitation into the vacuum bath, thereby favoring amplification over attenuation $\hat{\sigma}_-\hat{a}e^{2i\omega t}$. Incidentally, this very same dissipative process is the reason for the significant wastage of the work investment. This understanding is confirmed by looking explicitly at the heat transfer from the qubit to the thermal bath
\begin{equation}\label{eq:heat}
	\frac{\rmd \mathcal{Q}_\text{sys}}{\rmd t}=\text{Tr}\left[\hat{H}_\text{sys} \mathcal{L}_0 \left[\hat{\rho}\right]\right] =-\frac{\hbar \omega \gamma}{2}\left(1+\avg{\hat{\sigma}_z}\right)\xrightarrow{ss}-\frac{\hbar \omega \gamma}{2} ,
\end{equation}
which completes the thermodynamics analysis of the engine. As illustrated in Fig.\,\ref{fig:pow}, half of the work investment indeed flows to the qubit, and is entirely rejected as wasted heat once the system operates in the steady-state regime.

Our results thus imply that the use of a single generalized thermal bath to power the amplifier is indeed complying with the rules of thermodynamics, with the machine of Fig.\,\ref{fig:amp}(b) performing a conversion of work into work and heat. This insight suggests that a more efficient use of the work source may be available for the task at hand. For instance, the access to a resonant laser drive would allow to directly pump the cavity mode, thereby profiting from all the pump power without wasting any of it as heat.
\emph{Conclusion.--} We have identified the work cost of employing any generalized thermal bath. By mapping the dissipative coupling to the original thermal channel, our universal method allows to bring these out-of-equilibrium resources back to the realm of thermodynamics, where any ordering of the energy requires a work source.

Revisiting one of the oldest quantum heat machines aimed at amplifying light, we showed that the possibility to run the engine on a single bath does not contradict the laws of thermodynamics. We also found that a significant part of the work investment is rejected as heat. 

Our results highlight that generalized thermal baths can be useful when their work cost is not an issue. However, when aiming to reduce the amount of heat generated by an engine, a direct access to the work supply may allow for a more efficient use of energy.

\begin{acknowledgments}
We would like to thank C.~Bruder for discussions. Numerical simulations were performed using the QuantumOptics package in Julia~\cite{kramer18}.
This work was financially supported by the Swiss SNF and the NCCR Quantum Science and Technology.
\end{acknowledgments}

\bibliography{squeezeBib}

\end{document}